\documentclass[]{aa} %
\usepackage{graphics,txfonts,longtable}  
\usepackage{rotating}

\newcommand{\kms}{km\,s$^{-1}$}
\newcommand{\bz}{$\langle B_{\rm z} \rangle$}

\newcommand{\te}{$T_{\rm eff}$}

\newcommand{\lgg}{$\log g$}
\def\inv{\,{\sc INVERS10}}

\newcommand{\vs}{${\rm v}_{\mathrm{e}}\sin i$}

\def\i{\,{\sc i}} \def\ii{\,{\sc ii}} \def\iii{\,{\sc iii}}
\def\hd{HD\,24712}

\begin{document}

\title{Magnetic Doppler imaging of the roAp star HD\,24712%
\thanks{Based on observations made with the Nordic Optical Telescope, operated on the island of
La Palma jointly by Denmark, Finland, Iceland, Norway, and Sweden, in the Spanish Observatorio
del Roque de los Muchachos of the Instituto de Astrofisica de Canarias.}}
%\subtitle{Magnetic Doppler imaging of the roAp star HD\,24712}

\author{T. L\"uftinger\inst{1}, O. Kochukhov\inst{2}, T. Ryabchikova\inst{1,3}, N. Piskunov\inst{2}, 
 W.W. Weiss\inst{1}, \and I.~Ilyin\inst{4}
        }

\titlerunning{First magnetic Doppler images of an roAp star }
\authorrunning{T. L\"uftinger et al.}
\offprints{~\\ T. L\"uftinger at \email{lueftinger@astro.univie.ac.at}}

\institute{Institut f\"ur Astronomie, Universit\"at Wien, T\"urkenschanzstrasse 17, 1180 Wien, Austria
      \and Department of Physics and Astronomy, Uppsala University Box 515, SE-751 20 Uppsala, Sweden
      \and Institute of Astronomy, Russian Academy of Sciences, Pyatnitskaya 48, 119017 Moscow, Russia
      \and Astrophysikalisches Institut Potsdam, An der Sternwarte 16, D-14482 Potsdam, Germany
         }

\date{Received / Accepted }

\abstract {}  
{We present the first magnetic Doppler images of a rapidly oscillating Ap (roAp) star.}
{We deduce information about magnetic field
geometry and abundance distributions of a number of chemical elements on the 
surface of the hitherto best studied roAp star, \hd, using the magnetic Doppler
imaging (MDI) code, \inv, which allows us to reconstruct simultaneously and 
consistently the magnetic field geometry and elemental abundance 
distributions  on a stellar surface. 
For this purpose we analyse time series spectra obtained in Stokes I and V
parameters with the SOFIN polarimeter at the Nordic Optical Telescope and recover surface abundance structures of 
sixteen different chemical elements, respectively ions, including Mg, Ca, Sc, Ti, Cr, Fe, Co, Ni,
Y, La, Ce, Pr, Nd, Gd, Tb, and Dy. For the rare earth elements (REE) Pr and Nd
separate maps were obtained using lines of the first and the second ionization stage.}    
{We find and confirm a clear dipolar structure of the surface magnetic field 
and an unexpected correlation of elemental abundance with respect to this field: one group of
elements accumulates solely where the positive magnetic pole is visible,  
whereas the other group avoids this region and is enhanced where the magnetic equatorial region dominates 
the visible stellar surface.  
We also observe relative shifts of abundance
enhancement- or depletion regions between
the various elements exhibiting otherwise similar behaviour. }      
 {}             


\maketitle

\section{Introduction}
\label{introduction} 
Many physical processes important for stellar evolution 
are crucially influenced by stellar magnetic fields. Theoretical and observational frameworks 
tell us for instance that they influence microscopic diffusion of chemical elements within
stellar atmospheres, and they may also redistribute angular momentum or
give rise to enhanced hydrodynamical instabilities and thus modify mixing
properties in stellar interiors.     

Ap stars, representing about 
 5--10\%
of the upper main sequence stars,  
exhibit magnetic fields that appear to be
highly ordered, very stable and often very strong. 
Many Ap stars also show strong spectral line profile variations synchronized to stellar
rotation. This is attributed to 
oblique magnetic and rotation axes and to the presence of a non-uniform distribution of
chemical elements on their surface within the framework of the oblique rotator model (introduced
by Stibbs, \cite{stibbs1950}). With a few exceptions such inhomogeneities exist only in the
atmospheres of A stars with magnetic fields, demonstrating that these fields play a
crucial role in their formation and evolution. 

The spectra of Ap stars also exhibit a remarkable variety of sometimes unidentified spectral line
features. Ryabchikova~et~al. (\cite{Ryabchik04}) for example find
overabundances of some Fe peak elements, which may reach a few dex for Cr and
even higher for the REE, especially for their second ionization stage,  
while other chemical elements are found to be underabundant compared to the solar value. 
An important subgroup of the Ap stars, the rapidly oscillating Ap stars,
in addition exhibits high-overtone, low-degree, non-radial 
$p$-mode pulsations with periods of 6--21 minutes (Kurtz \& Martinez \cite{KM00}). Their observed pulsational amplitudes
 are modulated
with the magnetic field variations, exhibiting maximum amplitude at the phases of
highest magnetic field strength (Kurtz \cite{K82}). This indicates a close
connection between the magnetic field and the excitation mechanism of roAp 
pulsations. 
 
The chemical peculiarities mentioned are attributed to the selective diffusion
of ions under the competitive action of radiative acceleration and gravitational
settling (Michaud \cite{michaud1970}) 
possibly in combination with a weak, magnetically driven wind 
(see, e.g., Babel \cite{babel92}).

We focus our research on roAp stars since these objects exhibit a number of new, largely unexplored phenomena 
related to their pulsational variability while sharing all the properties of other Ap stars. The roAp stars
are the only main-sequence stars in which high-overtone p-mode pulsations are easily detected and pulsation 
modes can be identified. 
Furthermore, roAp stars represent the only group of pulsating stars for which magnetic and pulsation 
characteristics can well be constricted observationally. Hence, peculiar pulsating stars are 
key objects in the investigation of the atmospheric and internal structure of the
middle main sequence stars.

Very little is known as yet about the origin and structure of
magnetic fields and their connection and interaction with surface abundance
patches and pulsation. With the development of high-resolution spectropolarimeters it has become possible to 
extract the full amount of information about magnetic field and abundance distributions using 
Stokes parameter observations and applying magnetic Doppler imaging.

We have developed and tested a code for MDI  
(Piskunov\,\&\,Kochukhov \cite{PK02}; Kochukhov \& Piskunov \cite{KP02}) that
allows us to reconstruct
simultaneously and consistently the magnetic field vector and abundance distribution of various elements 
on a stellar surface by taking into account all relevant physics of polarized line formation. 

In deriving the surface abundances of numerous light-, iron-peak- and
rare earth elements by applying MDI, we get more insight into 
the relation between the magnetic fields, vertical and horizontal abundance characteristics and
pulsation, and thus into the atmospheric structure of Ap and roAp stars.
At present surface abundance structures have been derived for only one other
roAp star, HD\,83368 (Kochukhov et al. \cite{KDP04}). 
 
Our paper is composed in the following way: Sect.\,\ref{hd24712} is devoted to the
description of the target star, \hd, on which we comment in
Sect.\,\ref{observations} regarding our spectropolarimetric observations and the data reduction. 
Details about MDI, the magnetic field geometry of \hd, and
the surface abundance structures of individual elements can be found in
Sect.\,\ref{magdi} while Sect.\,\ref{discuss} is devoted to the discussion of our results.  

\section{\hd}
\label{hd24712} 
\hd\ (HD1217, DO Eri) is one of the best-studied roAp stars with light
(Wolff \& Morrison \cite{WoMo1973}), spectrum and magnetic variations that was discovered to pulsate
by Kurtz (\cite{K82}). Matthews et al. 
(\cite{MWWY88}) found radial velocity variations with an amplitude of
0.4$\pm0.05$\,\kms and the main photometric period of 6.14\,min. 

During an observing campaign with the Whole Earth Telescope (WET), Kurtz et al.
(\cite{KM2002}) detected a `missing' mode. The detailed investigations by 
Cunha et al. (\cite{cunha03} and \cite{cunha06}) suggest in this context, that the $p$-mode 
pulsations are strongly affected by the global stellar magnetic field.  

A study of the surface abundance inhomogeneity in HD~24712 based on the observed line profile
variations was performed by Ryabchikova~et~al. (\cite{Ryabchik00}). Surface element distribution was
schematically approximated as a set of circular spots whose position, radii
and the abundances inside and outside the spots
have been obtained with an iterative procedure. Ryabchikova et al.~(\cite{Ryabchik00}) found that
Pr\ii, \iii, Nd\ii, \iii, and Co are concentrated in one large spot near the magnetic pole, while
Fe is concentrated rather in 4 spots around the magnetic equator. 
They also found that centers of Nd, Pr and Co spots do not coincide with the
pole of the magnetic field geometry 
determined by Bagnulo et al. (1995), but are shifted 
by $\approx10\degr$ towards the magnetic equator. 
According to their analysis Pr-Nd and Co are overabundant by~$\approx$+1 dex in the spot relative to the
outside surface area and one needs 
$\approx$2 dex abundance differences between Pr and Nd in the first and the
second ionization stages. The maximum surface abundance gradient for Fe was 0.6 dex.   

We performed a detailed spectroscopic pulsation study of radial velocity variations in high
resolution and high time-resolved data 
(obtained with SOFIN at NOT and with UVES at the ESO VLT). Simultaneously we
performed high precision photometric observations 
from space with the MOST
(Microvariability and Oscillation of Stars) satellite
(Walker et al. \cite{WMK2003})  
During this process, we discovered a detailed pulsation picture of the star
(Ryabchikova et al. \cite{Ryabchik07}). 
We found maximum radial velocity amplitudes for rare
earth lines and in addition phase shifts progressively increasing from one element to another and within 
different groups of lines of the same element.
We propose that this pulsational behaviour is caused by the fact
that   
these elements are concentrated in a thin layer in the stellar atmosphere where pulsation reaches its maximum amplitude.

These observations and a possible correlation to theoretical
investigations made \hd\ an ideal candidate for spectroscopic and
spectropolarimetric observations with ground based equipment.

The basic physical parameters of \hd\ are listed in Table\,\ref{phy}.
\te\ and \lgg\ were applied as derived by
Ryabchikova et al. (\cite{Ryabchik97}). 
The atomic data used in our study was extracted from the Vienna Atomic Line Database 
(VALD, Piskunov et al. \cite{vald1}; Ryabchikova et al. \cite{vald2}; Kupka et al. \cite{vald3}).    

\begin{table}
\caption{Basic physical parameters of \hd.}
\label{phy}
\begin{center}
\begin{tabular}{ll}
\noalign{\smallskip}
\hline
\hline
\te           & $7250\pm 150\,K $ \\
\lgg          & $4.2\pm 0.1$ \\
$M$             & $1.63\,M_{\odot}$ \\ 
$L/L_{\odot}$ & $0.91\pm 0.04$ \\
$P_{\rm rot}(\mathrm{d})$     & $12.45877\pm(16) $ \\
$P_{\rm puls}$    & $\simeq 6.15\,\mathrm{min}  $ \\
\vs & 5.6 \kms\\
\hline
\end{tabular}
\end{center}
\end{table}% 

\section{Observations and data reduction}
\label{observations} 
Spectropolarimetric observations of \hd\ were carried out in 2003, between Oct.
29 and Nov. 18, using the high-resolution \'echelle spectrograph SOFIN 
attached to the Cassegrain focus of the 2.56\,m Nordic Optical Telescope (NOT), La Palma, Spain. 
The spectrograph is equipped with three different cameras, and to obtain
observations in the spectropolarimetric 
mode, the second camera was used. For MDI we used 49 different spectral lines from seven spectral orders, each covering about
40\,\AA\ to 50\,\AA\ between 4000 and 
6000\,\AA. 

The polarized spectra were obtained with a Stokesmeter, consisting of a calcite
plate used as a beam splitter and an achromatic rotating quarter-wave plate.
The position of the quarter-wave plate is controlled by 
a stepping motor. Usually a sequence of four subexposures is processed in order to obtain accurate circular polarization 
measurements, where each of the beams is exposed twice, with the quarter-wave plate rotated by $90\degr$ after the first 
and before the last exposure. This way it is possible to reduce instrumental effects
to a minimum, because in the images taken with the quarter-wave 
plate rotated by $90\degr$, possible instrumental signatures change sign 
and cancel each other out during the averaging of the two exposures. 

For data reduction, the 4A software package written in C (Ilyin \cite{II00}) was
used. All standard procedures like bias subtraction, flat field correction, subtraction of the scattered light, 
weighted extraction of the orders, 
and bad pixel (cosmic ray) corrections were included. ThAr exposures obtained before and after each observing night 
were used to perform wavelength calibration and to test for possible spurious
instrumental polarization, caused for instance by bending of the 
spectrograph, which is directly mounted on the telescope, different positions of
the star on the slit or temporal 
variations of the seeing. S/N ratios for the observed spectra typically reached
500--600 per pixel at $\lambda$\,=\,5000~\AA, except for one spectrum which has S/N\,$\approx$\,300. Reduced spectra have
a resolving power of $\approx$\,40\,000.

In total we obtained observations at 13 rotational phases. 
Each pair of the Stokes $I$ and $V$ spectra was calculated from a sequence of polarimetric
exposures obtained over 35 min. This is sufficiently long to average out
variations due to the 6.15~min pulsation period of \hd.

Rotation phases of \hd\ (see Table~\ref{stokes}) were calculated according to the ephemeris and
rotation period derived by Ryabchikova et al. (\cite{Ryabchik05}): \\\\
\hspace{3mm}HJD(\bz$_{max})=2453235.18(40)+12.45877(16)$\,d.

\begin{table}[t]
\caption{Journal of spectropolarimetric observations of \hd.}
\label{stokes}
\begin{center}
\begin{tabular}{ccc}
\hline
\hline
   HJD      & Rotational phases & $S/N$\\
(245 0000+) &                   &      \\
\hline	   
2941.6516 &  0.4396  & 290 \\ %& 0.4400
2943.6341 &  0.6001  & 550 \\ %& 0.5991
2945.5758 &  0.7562  & 520 \\ %& 0.7550
2946.5973 &  0.8380  & 510 \\ %& 0.8370
2947.6067 &  0.9171  & 570 \\ %& 0.9180
2948.6514 &  0.0026  & 660 \\ %& 0.0018
2952.6267 &  0.3207  & 560 \\ %& 0.3209
2953.6572 &  0.4040  & 520 \\ %& 0.4036
2954.6316 &  0.4824  & 550 \\ %& 0.4818
2955.6350 &  0.5620  & 570 \\ %& 0.5624
2956.5954 &  0.6390  & 620 \\ %& 0.6395
2957.5958 &  0.7204  & 560 \\ %& 0.7198
2961.6160 &  0.0414  & 500 \\ %& 0.0424
\hline
\end{tabular}
\end{center}
\end{table}

\section{Magnetic Doppler imaging (MDI)} 
\label{magdi}

Doppler imaging (DI) has become a very successful tool to gain information on
inhomogeneities on the surface of a rotating star caused by changing abundance and/or
magnetic field structure.
The first method for reconstruction of stellar magnetic field geometries from Stokes 
V and I spectropolarimetry was developed by Piskunov (1985) and applied to Ap stars by 
Glagolevskii et al. (1985). A similar technique has been very successfully applied to
mapping magnetic 
field topologies in late-type active stars by Donati \& Collier Cameron~(1997),
Hussain et al.~(2001), and Donati~et~al.~(2006).

By inverting time series of spectropolarimetric observations of rotating
stars based on regularized image-reconstruction procedures implemented in \inv\ developed by Piskunov
\& Kochukhov (\cite{PK02}), it became possible to invert simultaneously the information of the 
rotationally modulated Stokes profiles into elemental abundance as well as the three vector components of 
the stellar magnetic field. 
In their papers\,{\sc I} and {\sc II}, Piskunov\,\&\,Kochukhov (\cite{PK02}) and
Kochukhov\,\&\,Piskunov (\cite{KP02}) describe in detail the concept of MDI, the basic numerical
techniques to efficiently solve the magnetic radiative transfer, the
surface integration of the resulting Stokes profiles and the application of the
regularization concept. Applications of \inv\ are presented in Kochukhov et al.
(\cite{KPI02,KBW04}) and Folsom et al. (\cite{FWK08}).

In case of \hd\ we only have access
to two Stokes parameters (I and V). For this reason we had to resort to
multipolar regularization to obtain a solution free 
from undesirable cross-talks between different field components. 
This is a reasonable prior model of the Ap-star magnetic field geometry. Numerous previous studies (e.g. Landstreet
\& Mathys \cite{LM00}; Bagnulo et al. \cite{BLL02}) showed that rotational modulation of the integral quantities
derived from the circular polarization  
spectra of Ap stars are well-explained by dipolar or quadrupolar-like fields with  
some additional influence of abundance inhomogeneities. 
Multipolar regularization (see Piskunov\,\&\,Kochukhov
\cite{PK02}) has been successfully used for the mapping of the Ap stars
$\alpha^2$\,CVn (Kochukhov et al. \cite{KPI02}) and HD\,72106 (Folsom et al.
\cite{FWK08}).
This method introduces external 
constraints similar to the multipolar field parameterization proposed by Bagnulo et al.
(\cite{BLL99}) and leads \inv\ to solutions close 
to a general second-order multipolar expansion. Numerical experiments 
(Kochukhov\,\&\,Piskunov \cite{KP02})
demonstrated that this technique provides reliable imaging
of stars with quasi-dipolar or quadrupolar magnetic fields. However, our previous experience with the
MDI shows that using multipolar regularization may prevent us from reconstructing the fine structure of
magnetic field by forcing it closer to a low-order multipolar geometry.
The alternative, Tikhonov 
regularization, which does not introduce assumptions about the global geometry of the stellar surface
structures, was only applied for reconstruction of chemical abundance.
The values of two regularization parameters were adjusted to ensure that at the convergence
the total contribution of the regularization terms was roughly 20--80\% of the 
the sum of weighted squared differences between observations and synthetic spectra. We have verified
that a variation of regularization parameters by factors of 2--5 leaves little imprint on the final
solution.

The additional information content of the Stokes profiles compared to the data
used for conventional Doppler Imaging makes it possible to also
map stars with rotational velocities as low as that of \hd\ with
\vs=5.6\,\kms\ (derived by Ryabchikova et al. \cite{Ryabchik97}). 
Nevertheless we would like to state here that due to this very
low \vs, the uncertainties of the latitudes of the abundance enhancement and depletion regions are larger  
than for stars with \vs\ higher than $\simeq$\,10~\kms. 
On the other hand, the longitude position of the surface features is mainly constrained by
the phase coverage, which is fairly good for our observations.
Information contained in the considerable rotational modulation of the line intensities due to
changing visibility of abundance spots on the surface of \hd\ allows us to infer
the longitudinal
position of large-scale abundance structures with sufficient precision.
Numerical experiments demonstrate that longitudinal shifts of major abundance features by 
$\sim$\,10\degr\ (few 0.01 in phase units) can be readily detected with our observational data.

\begin{figure*}[htbp]
\vspace{7.1mm}
\begin{center}
\includegraphics[width=130mm]{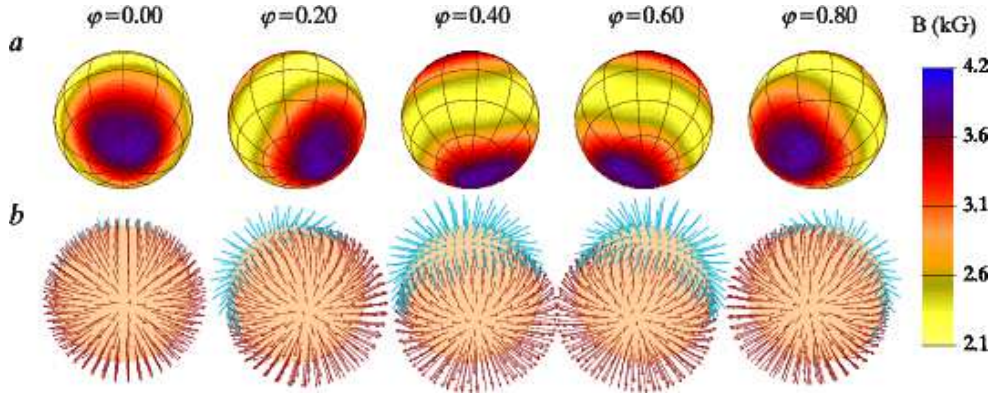}
\vspace{3.5mm}
\caption{Mapping of the distribution of magnetic field strength (a) and field orientation (b) on the surface
of the roAp star \hd, as derived with modelling the Stokes I and V parameter stellar
observations. The results reveal a dominant dipolar magnetic field geometry on the surface of this
star with the field strength varying between 2.1\,kG and 4.2\,kG.}
\label{mf}
\end{center}
\end{figure*}

\begin{figure*}[htbp]
\begin{center}
\includegraphics[width=130mm]{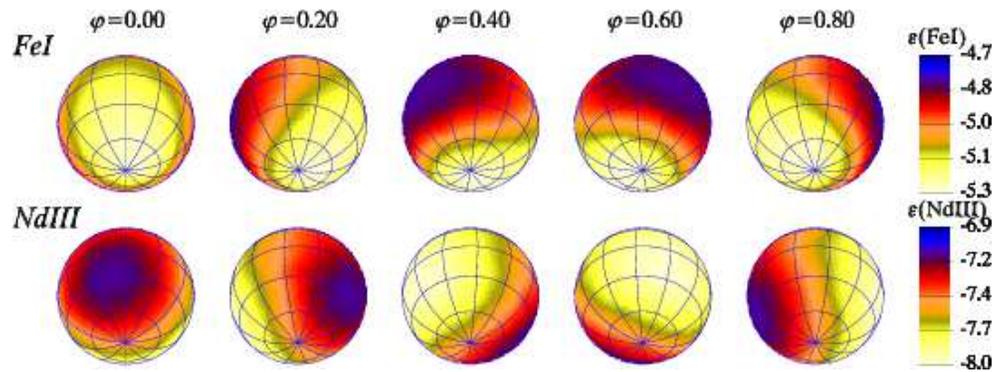}
\caption{The abundance distributions of Fe  and Nd on the surface of \hd. The upper panel shows 
the surface map of the Fe abundance and the lower panel presents the
distribution of Nd derived from Nd\iii\ lines.  
These maps were derived using the Stokes I and V parameter spectra as presented
in Fig.\,\ref{profFeNd}. In this and other similar plots we use the $\varepsilon(Xi)$ to denote abundance, in the 
$\log(N_{\rm X}/N_{\rm tot})$ units of the element ``X'', derived from lines
belonging to the ionization stage ``i''.} 
% and assuming multipolar field geometry First self-consistent mapping of an abundance distribution and stellar 
\label{abnFeNdIII} 
\end{center}
\end{figure*}

\subsection{Magnetic field geometry} 
\label{mfg} 

Determining the geometry of the magnetic field on the surface of \hd\ was
performed choosing seven different Fe\i\ and five different Nd\iii\ lines, which are listed in Table\,\ref{lines}. These lines were chosen 
because their Land\'e factors and Zeeman splitting patterns yield strong polarization signatures, they are not
affected by severe blending, and they show clear variations in the strength of the Stokes I profiles
due to abundance inhomogeneities and, to a lesser extent, due to variable Zeeman effect.
Furthermore, the
lines of Fe and Nd exhibit an opposite variation with rotation phase. Thus, using these two elements
simultaneously in a self-consistent abundance and magnetic inversion yields a more robust
reconstruction of the stellar magnetic field topology.

\begin{figure*}[htbp]
\begin{center}
\vspace{10.7mm}
\includegraphics[width=0.8\textwidth]{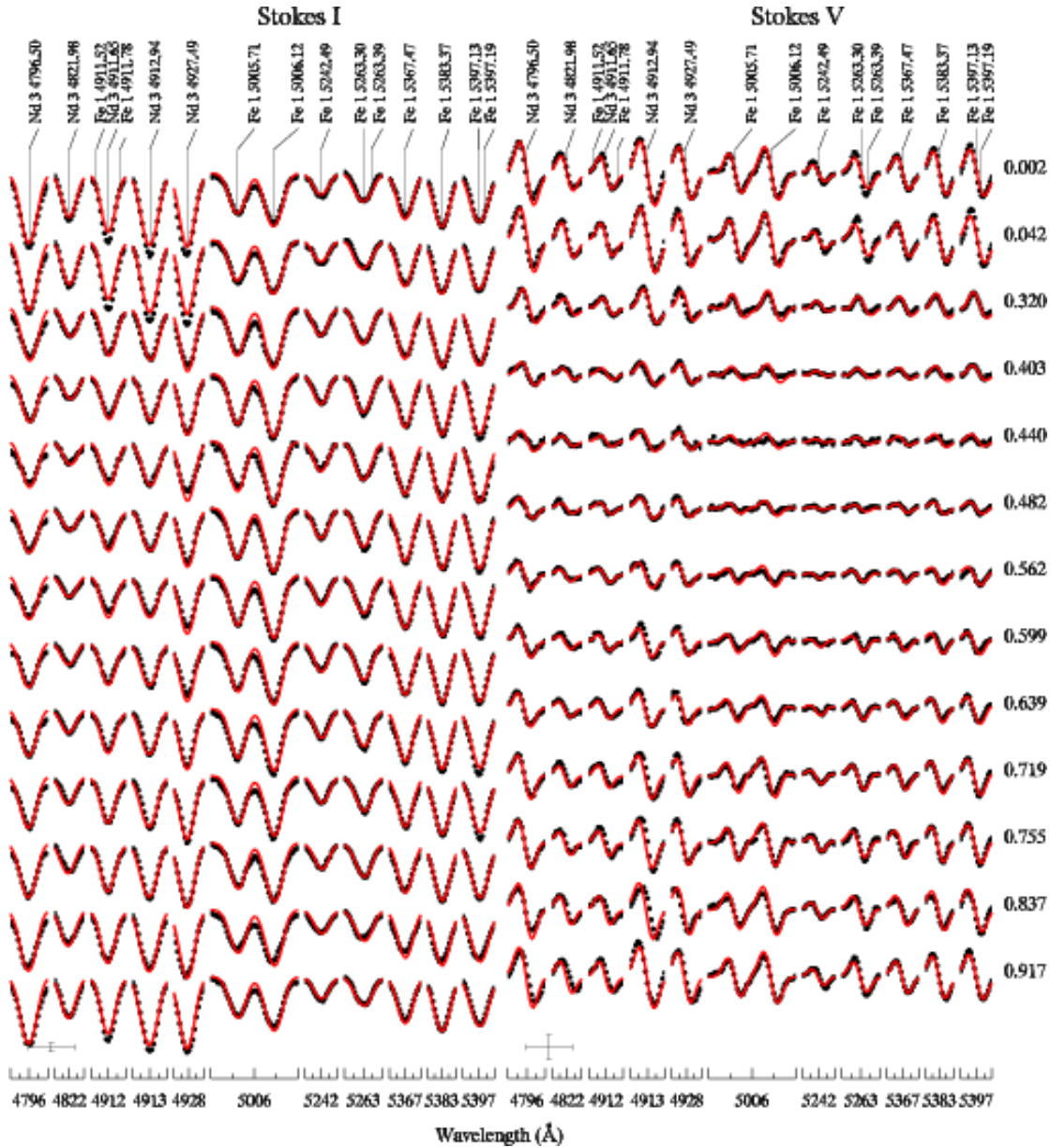}
\vspace{4.4mm}
\caption{Comparison of the observed (dots) and calculated (lines) Stokes I and V
profiles. Spectra are shown with increasing phase value (listed in Table
\ref{stokes}) from top to bottom. The bars at the lower left and middle show the
vertical (5\%) and horizontal (0.5 \AA) scale.}
\label{profFeNd}
\end{center}
\end{figure*}

Important parameters needed as input for the MDI procedure are the inclination and the
azimuth angle of the stellar rotational axis, $i$ and $\Theta$. It should be
mentioned that when
mapping with Stokes I and V components, the azimuth angle is irrelevant and we cannot
distinguish between $i$ and $180\degr-i$. However, we adopted $i > 90\degr$ to be consistent with the analysis
of Bagnulo et al. (\cite{Bagn95}), who determined both angles based on the traditional circular-
and on broadband linear polarization observations and found
$i=137\degr$ and $\Theta=4\degr$.
Adopting these values, 
a surface magnetic map very close to a dipolar geometry 
yielded the best 
fit to the observed line profiles. The magnetic field strength 
varies between $2.1$\,kG and $4.2$\,kG. In Figs. \ref{mf} and
\ref{profFeNd} the observed line profiles of the chosen Fe\i\ and Nd\iii\
lines (dots), the corresponding fit reached during the inversions  
(lines) and the resulting magnetic field geometry are presented. The near dipolar
structure agrees very well with the dipolar model proposed by
Bagnulo et al. (\cite{Bagn95}), only differing in the polar field strength, where
they obtain $B_{\rm p} = 3.9$\,kG, compared to $4.2$\,kG derived from magnetic
Doppler Imaging.

\subsection{Fe and Nd abundance distribution}
\label{sec:FeNd}
The surface abundances of Fe and Nd mapped simultaneously with the
stellar magnetic field geometry are shown in Fig.\,\ref{abnFeNdIII}. As mentioned above, 
these distributions were calculated using 
seven Fe\i\ and five Nd\iii\ lines listed in Table\,\ref{lines}. 
One can easily see that abundances of both elements are quite globally
structured. Fe varies between $-5.3$ and $-4.7\,$dex in the $\log(N/N_{\rm tot})$, in its maximum,
still slightly underabundant compared to the solar value (-4.59\,dex). Nd abundance being derived from Nd\iii\ lines
is very high and it varies between $-8.0$ and $-6.9\,$dex (compared to $-10.59\,$dex solar value).
As we see further in Sect.~\ref{ree} Nd abundance distribution from Nd\ii\ lines is similar in shape but by 1.5 dex smaller 
in magnitude following the known REE anomaly observed in all roAp stars (Ryabchikova et al. \cite{Ryabchik01}, \cite{Ryabchik04}). 
Fe and Nd seem to be perfectly anticorrelated: 
Fe is accumulated where Nd is depleted, and minimum Fe abundance 
can be found where Nd is at its maximum.  

Comparing the patterns of these two elements to the derived magnetic field
geometry presented in Fig.\,\ref{mf}, we find that the Fe abundance enhancement region is associated with
the area where the magnetic equator dominates the visible stellar hemisphere,
around $\phi\simeq 0.5$, whereas the Nd map  
exhibits its area of maximum abundance around the phase where the positive magnetic pole is oriented 
towards the observer. Due to the orientation of the star we are not able to directly observe the negative 
magnetic pole.		 

It is surprising to see that, contrary to our expectations, both elements are not accumulated or depleted on 
\emph{both} magnetic poles or around the \emph{magnetic equator}. 
Furthermore, the center of maximum or minimum abundance, respectively, seems to be shifted in latitude 
and/or longitude with respect to the magnetic polar regions, not only for Nd, but also for several other elements presented 
in Sect.\,\ref{abundn}.  

\begin{figure*}[htbp]
\begin{center}
\includegraphics[width=0.7\textwidth]{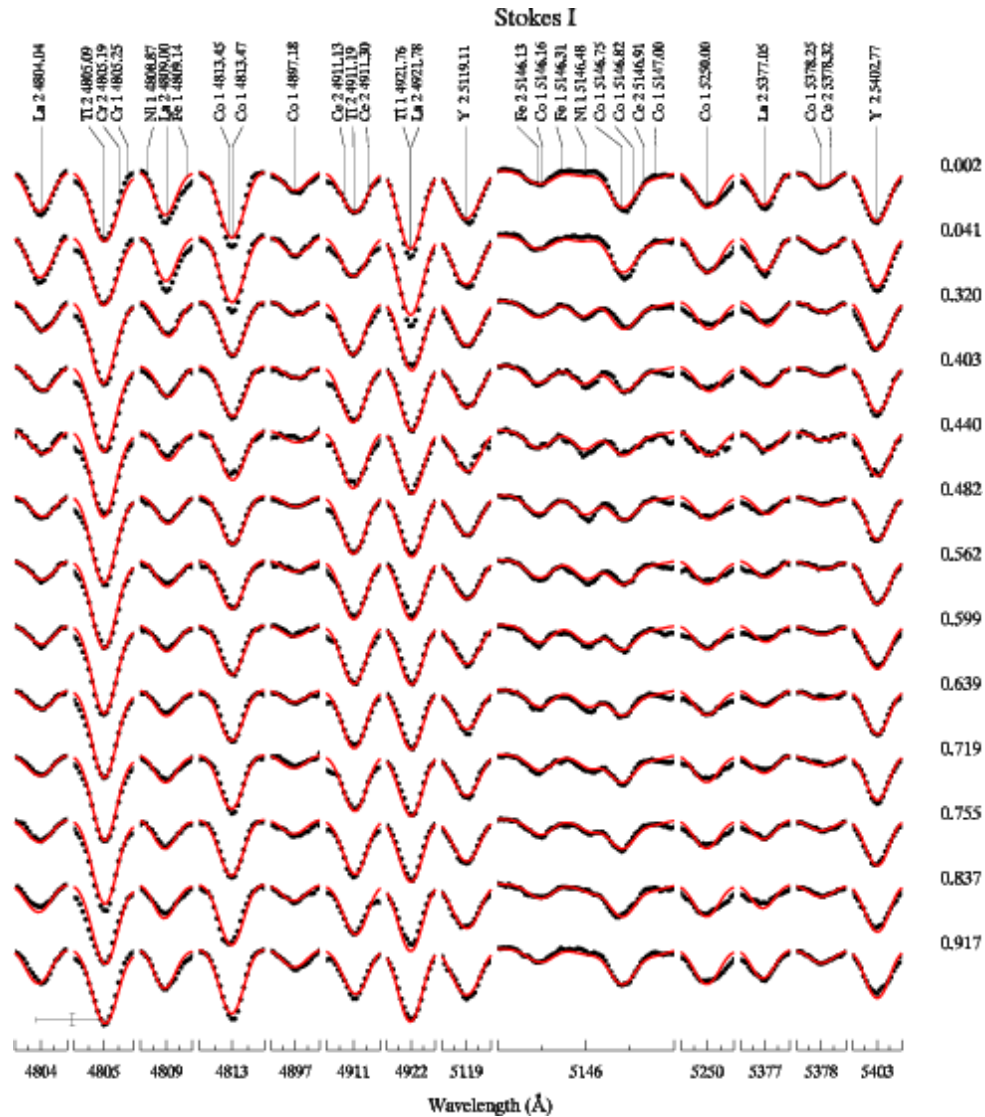}
\caption{Comparison of the observed (dots) and calculated (lines) Stokes I and V profiles used 
for mapping of Ti, Co, Y and La. Spectra are plot with increasing phase value (listed in Table
\ref{stokes}) from top to
bottom. The bars at the lower left and middle again show the vertical (5\%) and
horizontal (0.5 \AA) scale.
Note the total disappearance and reappearance of the Ni\i\ line at 5146 \AA.}
\label{prof:CoI}
\end{center}
\end{figure*}

\begin{table*}
\caption{Table of elements and  spectral lines used for mapping fromleft to
right: 
 element identification, central wavelength,  
the log{\it gf} values of these lines, abundance interval and the range in dex, and the solar value
as a comparison presented in columns from left to right. }  
%Atomic parameters used
%in our study were extracted from the Vienna Atomic Line Database (VALD Piskunov~et al. 1995, Ryabchikova et al. 1999, Kupka et al. 1999).}
\label{lines}
\begin{center}
%\begin{tiny}
%\begin{scriptsize}
\begin{tabular}{|l|r|c|c|c|c|} 
\hline
Element & $\lambda$ (\AA) & log\it gf & min, max & abundance interval & $\odot$  \\
        &                 &           & $\log(N/N_{\rm tot})$ & $\log(N/N_{\rm tot})$   &  $\log(N/N_{\rm tot})$ \\
\hline 
\hline 
Mg \i\      & 5528.405 & -0.620 & $-6.0, -4.5$  & 1.5 & -4.51 \\
\hline 
Ca \ii\     & 5021.138 & -1.456 & $-5.2, -5.0$  & 0.2& -5.73 \\
\hline 
Sc \ii\     & 5526.790 &  0.024 & $-10.1, -8.9$ & 1.2& -8.99 \\
\hline 
Ti \ii\     & 4805.085 & -0.960 & $-7.8, -7.1$  & 0.7& -7.14 \\
                   & 4911.195 & -0.610 &  & &     \\
%                   & 5016.161 & -0.574 &  & &     \\
\hline 
Cr \ii\     & 5510.702 & -2.614 & $-6.1, -5.9$  & 0.2& -6.40 \\
\hline 
Fe \i\	    & 5005.711 & -0.180 & $-5.3, -4.7$  & 0.6 & -4.59 \\
         	   & 5006.117 & -0.638 &   & &     \\
        	   & 5242.490 & -0.967 &   & &     \\
        	   & 5263.300 & -0.879 &   & &     \\
        	   & 5263.394 & -2.774 &   & &     \\
        	   & 5367.470 &  0.443 &   & &     \\
        	   & 5383.370 &  0.645 &   & &     \\
	           & 5397.130 & -1.993 &   & &     \\
	           & 5397.193 & -0.342 &   & &     \\

\hline 
Co \i\      & 4813.449 & -2.121 & $-6.4, -5.8 $  & 0.6 & -7.12  \\
               	   & 4813.467 &  0.050 &  & &     \\
               	   & 4897.179 & -0.764 &  & &     \\
	           & 5146.161 & -3.660 &  & &     \\
	           & 5146.749 & -0.320 &  & &     \\
	           & 5146.823 & -2.393 &  & &     \\
	           & 5146.998 & -1.286 &  & &     \\
	           & 5250.000 &  0.320 &  & &     \\
	           & 5378.247 & -0.425 &  & &     \\
%	           & 5393.739 & -0.326 &  & &     \\
\hline 
Ni \i\      & 5035.357 & 0.290 & $-7.6, -6.3$  & 1.3& -5.81 \\
	           & 5146.480 & -0.060 &  & &	 \\
\hline 
Y \ii\      & 5119.112 & -1.360 & $-9.0, -8.5$  & 0.5& -9.83  \\
                   & 5402.774 & -0.510 &  & &	 \\
\hline 
La \ii\     & 4804.039 & -1.494 & $-9.8, -8.8$  & 1.0& -10.91 \\
                   & 4808.996 & -1.494 &  & &     \\
                   & 4921.776 & -0.699 &  & &     \\
                   & 5377.052 & -0.316 &  & &     \\
\hline 
Ce \ii\     & 5117.169 & -0.050 & $-9.3, -8.9$  & 0.4&  -10.46   \\
	           & 5273.899 & -1.210 &   & &     \\
	           & 5274.229 &  0.150 &   & &     \\

\hline 
Pr \ii\     & 5135.140 & -0.131 & $-10.0, -9.6$ & 0.4& -11.33 \\
Pr \iii\    & 4910.823 & -1.954 & $ -8.4, -7.7$ & 0.7&  \\
                   & 4929.115 & -2.068 &   & &     \\
\hline 
Nd \ii\     & 4811.342 & -1.015& $-9.3, -8.5$  & 0.8& -10.59 \\
                   & 5033.507 & -0.470 &   & &     \\
                   & 5132.328 & -0.710 &   & &     \\
                   & 5143.337 & -1.570 &   & &     \\
                   & 5276.869 & -0.393 &   & &     \\
                   & 5361.467 & -0.482 &   & &     \\
                   & 5385.888 & -0.860 &   & &     \\
                   & 5399.099 & -1.419 &   & &     \\
Nd \iii\    & 4796.499 & -1.657& $-8.0, -6.9$  & 1.1 & -10.59 \\
                   & 4821.980 & -2.405 &  & &     \\
                   & 4911.653 & -1.826 &  & &     \\
                   & 4912.943 & -1.789 &  & &     \\
                   & 4927.488 & -0.850 &  & &     \\
\hline 
Gd \ii\     & 5500.419 & -1.333& $-8.8, -8.5$  & 0.3 & -10.92 \\
\hline 
Tb \iii\    & 5505.409 & -0.815 & $-9.1, -7.7$  & 1.4& -11.76 \\
\hline 
Dy \ii\     & 4923.167 & -2.384 & $-8.8, -8.3$  & 0.5 & -10.90 \\
\hline
\end{tabular}
%\end{tiny}
\end{center}
\end{table*}

\begin{figure*}[htbp]
\begin{center}
\vspace{8.0mm}
\includegraphics[width=130mm]{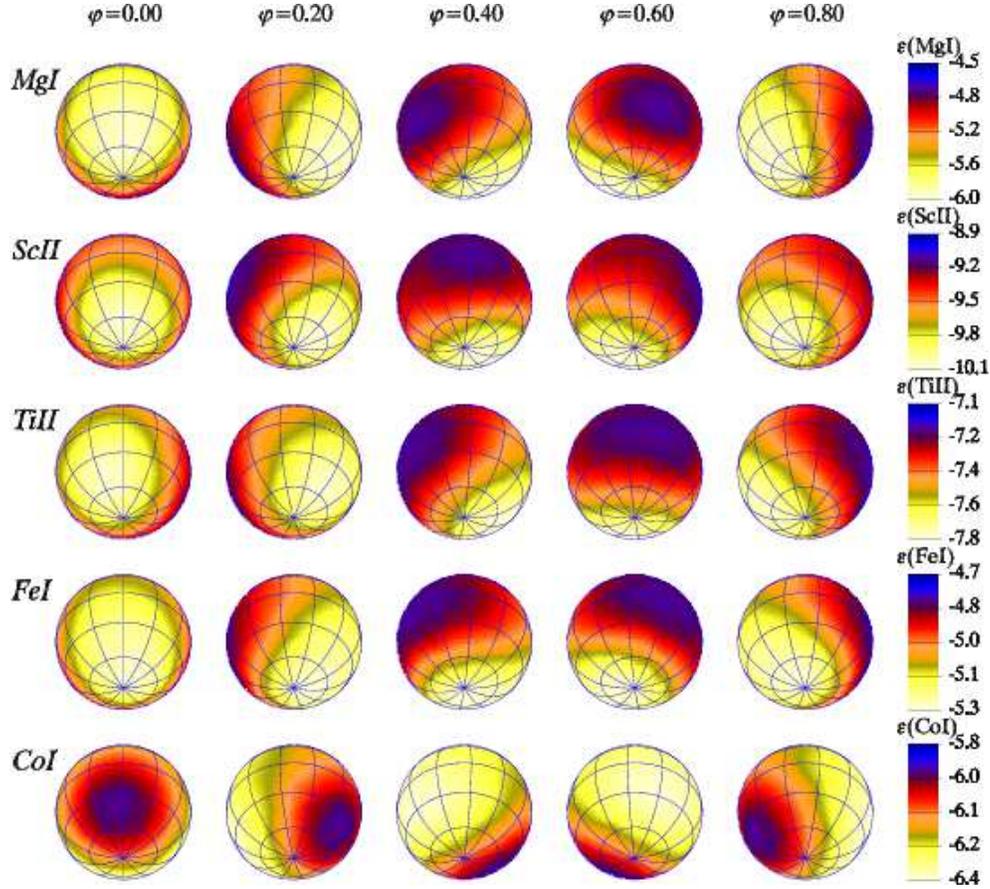}
\caption{Abundance distribution of Mg, Sc, Ti, Fe, and
Co on the surface of \hd\ obtained from the lines listed in
Table~\ref{lines}.}
\label{MgIScIITiIIFeICoI}
\end{center}
\end{figure*}

\begin{figure*}[htbp]
\begin{center}
\vspace{8mm}
\includegraphics[width=130mm]{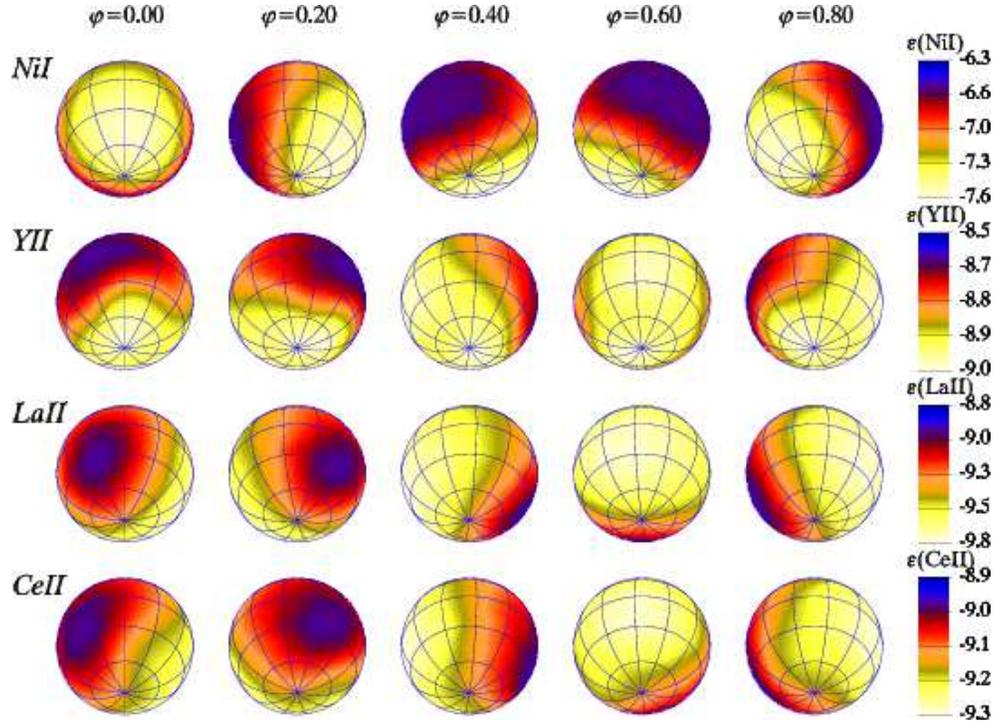}
\caption{Surface abundance distribution of Ni, Y, La, and
Ce (from top to bottom) on the surface of \hd. The maps were obtained
taking into account the lines listed in Table~\ref{lines}.}
\label{NiIYIILaIICeII}
\end{center}
\end{figure*}

\begin{figure*}[htbp]
\begin{center}
\vspace{8.1mm}
\includegraphics[width=130mm]{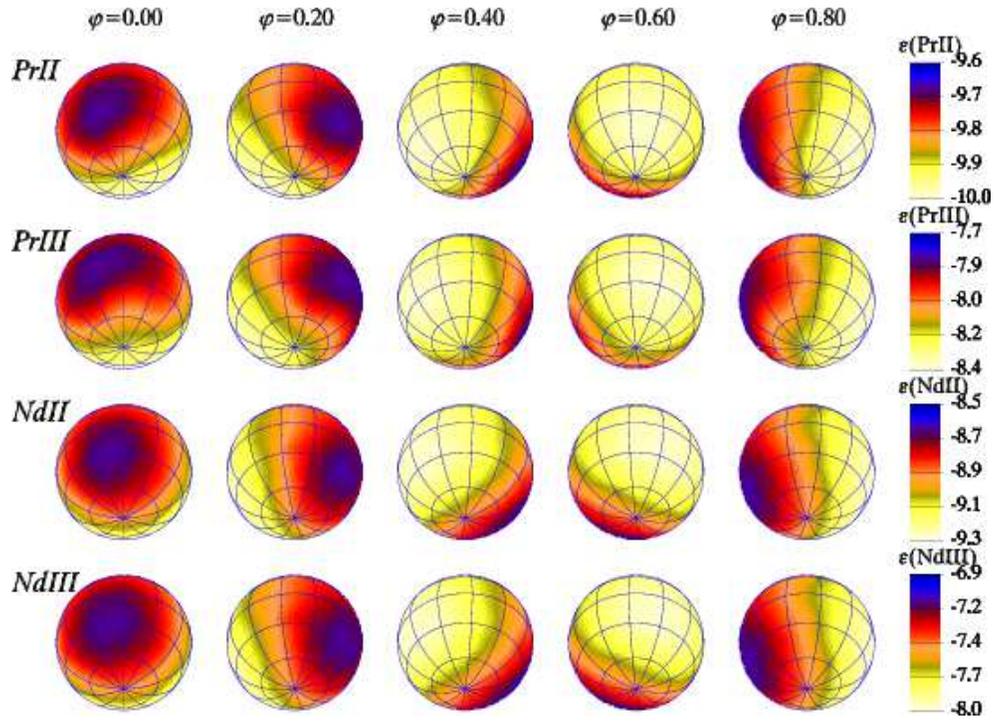}
\caption{Elemental abundance distribution on the surface of \hd. Maps of
Pr and Nd derived separately from the lines of the first and second ionization
stages of both elements are presented.}
\label{PrIIPrIIINdIINdIII}
\end{center}
\end{figure*}
 
\begin{figure*}[htbp]
\begin{center} 
\vspace{8mm}
\includegraphics[width=130mm]{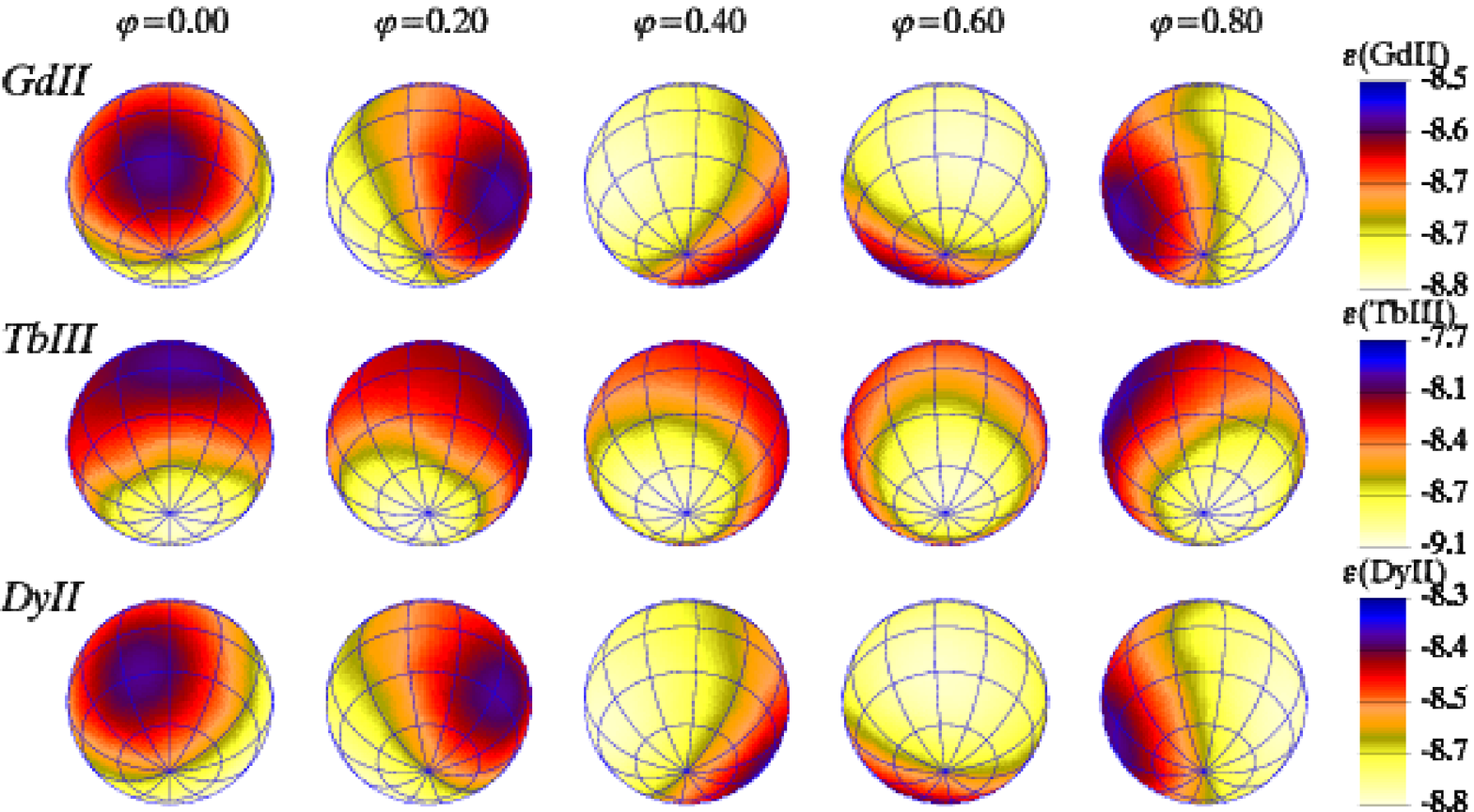}
\caption{Surface abundance distributions of Gd, Tb, and Dy 
derived from the lines listed in Table~\ref{lines}.}
\label{GdIITbIIIDyII}
\end{center}
\end{figure*}
 
\subsection{Surface abundances of individual elements}
\label{abundn}
After having derived the magnetic field geometry of \hd\ using the Fe\i\ and
Nd\iii\ lines, we were able to recover abundances of sixteen additional ions applying the
\inv\ code and the previously determined fixed magnetic map. 

\subsubsection{Magnesium, calcium, and scandium}
\label{MgCaSc}
The surface abundance distribution of \emph{magnesium} was modelled using one
Mg\i\ line at 5528\,\AA\ 
with \inv, and we found a remarkable 
variation between $-6.0$ and $-4.5$\,dex. \emph{Mg} is, as Sc,
Ti, Fe, and Ni depleted (underabundant by up to
1.5\,dex compared to the solar value) at the phase where the positive magnetic
pole dominates the visible stellar surface and shows a
single, huge overabundance region around the magnetic equatorial zone, indicating that the
center of maximum abundance does not coincide with the (suspected) negative magnetic
pole.  
 
For \emph{calcium} as well as for \emph{chromium} (see Section \ref{Fe-peak}) 
we found only a marginal abundance variation not exceeding 0.2 dex and consequently do not 
consider our mapping results as significant.

\emph{Scandium} exhibits a variation of 1.2\,dex over the stellar surface, 
between $-8.9$ and $-10.1$\,dex, also accumulated where the magnetic equatorial region dominates 
the visible stellar surface, 
whereby the central region of the depleted area occurs at similar longitude as that of 
Mg and Fe and the central region of maximum abundance precedes that of Mg and Fe
by about 60\degr. 
The region of enhanced \emph{Sc} abundance seems to occupy a larger fraction on 
the surface of \hd\ than numerous other elements. At its maximum abundance
phase, the \emph{Sc} 
value corresponds to the solar one of 8.99\,dex.

\subsubsection{Iron-peak elements} 
\label{Fe-peak}
Two lines of ionized \emph{titanium} were used to recover the surface abundance
structure of this element (Fig.~\ref{prof:CoI}). As Mg and Sc, it shows a tendency to
accumulate near the magnetic equatorial region, exhibiting near solar concentration,
and the abundance decreases to $-7.8$\,dex where the positive magnetic pole is visible. 

For \emph{chromium} only one line around 5510\,\AA\ could be used as input for
\inv\, and we obtained just a marginal (insignificant as for \emph{calcium}, see Section
\ref{Fe-peak}) variation between $-5.9$ and $-6.1$\,dex. 

\emph{Cobalt}, as Y and all the REE, is clearly accumulated on the surface
region visible at the phase
where the positive magnetic pole dominates, and less abundant around the magnetic equatorial region, varying between
$-5.8$ to $-6.4$\,dex. Co is significantly overabundant with respect to the Sun
($-7.12$\,dex) over the whole stellar surface. The element was mapped taking into account six Co\i\ lines shown
in Fig.~\ref{prof:CoI} and listed in Table\,\ref{lines}.    

A very interesting example of enormous line profile variation has been
discovered for the \emph{nickel} lines,
Ni\i\ at 5035\,\AA, and 5146\,\AA. The surface map derived from the single
unblended line at 5035\,\AA\ 
reveals a huge, high 
contrast region of overabundance at phase 0.5 and a region of underabundance
down to $-7.6$ dex at the phase where the positive magnetic pole dominates 
the visible stellar surface. We noticed that the Ni\i~$\lambda$5146.48\,\AA\ line totally disappears at this rotational phase.
In order to check the abundance derived from the single 5035\,\AA\ line and to be able to use the
blend of Ni with Fe, Co, and Ce around 5146\,\AA,  we used the abundance pattern obtained from
the Ni\i\ 5035 line to map Co and Ce from the blend mentioned above and got a very good
agreement of observed and calculated profiles. The fit to the line profiles
reached from this inversion is presented in Fig.~\ref{prof:CoI}.		       
Ni distribution, in longitude and in latitude, is very similar to that of Fe, and abundance
varies from $-7.6$ to $-6.3$\,dex, being significantly underabundant compared to the solar value of $-5.81$\,dex. 

\subsubsection{Yttrium and the rare-earth elements}
\label{ree}
\emph{Yttrium} varies from $-9.0$ to $-8.5$\,dex
(overabundant up to 1.3\,dex compared to the solar value of $-9.83$) and exhibits a structure closer to that of the REE, 
with the center of abundance maximum shifted to the northernmost part of the
visible stellar hemisphere. Yttrium reaches its maximum and 
minimum, respectively, slightly after that of Nd. 
As presented in Figs.~\ref{NiIYIILaIICeII}, \ref{PrIIPrIIINdIINdIII}, and \ref{GdIITbIIIDyII}, all the REE we mapped 
are enhanced 
around the phase where the positive magnetic pole is visible and depleted where the
magnetic equatorial region dominates the surface. For REE {\it depletion} means the minimum
 of the observed surface
abundance which usually
exceeds the solar photospheric abundance.    
We obtained abundance distributions
of La, Ce, Pr, Nd, Gd, Tb, and Dy. For Pr and Nd, abundance distributions were calculated 
from the separate analysis of the lines of the singly and doubly ionized element. 

\emph{Lanthanum} is significantly overabundant compared to the solar
value ($-10.91$) and varies from $-9.8$ to $-8.8$~dex. It
covers stellar latitudes very similar to those of Co, Nd, and Gd, although 
due to the very low \vs\ of the star, the latitudes 
of the abundance enhancement and depletion regions are affected by larger uncertainties 
than for stars with \vs\ higher than $\simeq$ 10\kms.
The center of maximum abundance slightly lags behind that of Nd.

Variation of Y and La lines is illustrated in Fig.~\ref{prof:CoI}.

The \emph{cerium} surface map was reconstructed using two Ce\ii\
lines, also exhibiting the typical 
overabundance of REE in roAp stars of more than a dex, but varying only moderately over the surface 
between $-9.3$ and $-8.9$~dex. 

Two independent maps of \emph{praseodymium} were derived using the lines of the 
singly and doubly ionized element. It is caused by the known REE anomaly
observed in roAp stars, where average abundance derived
from these lines differs by 1.5 -- 2.0 dex (Ryabchikova et al.
\cite{Ryabchik04}), and  
it is not possible therefore to fit line profiles of 
both ionization stages within the same inversion. Pr\ii~$\lambda$5136\,\AA\ and 
two spectral lines of  Pr\iii\ around 4910\,\AA\ and 4929\,\AA\ were used for inversions. Pr varies moderately between $-10.0$ and $-9.6$\,dex
being derived from one Pr\ii\ line, while higher abundances but with the same distribution 
shape are obtained from Pr\iii\ lines. 
The Pr spot seems to be shifted in latitude relative to the Nd enhancement region.
 
Abundances for \emph{neodymium} were also derived separately from singly and doubly
ionized lines due to the large abundance difference between the two 
ionization stages. The Nd surface map was derived from eight lines of Nd\ii\ listed in Table\,\ref{lines} 
and exhibits the same distribution 
as that obtained from Nd\iii\ lines (see Sect.\ref{sec:FeNd}). Nd abundance
varies from $-9.3$ to $-8.5$\,dex. As for the Pr case, the abundance difference
derived from Nd\ii\ and Nd\iii\ lines reaches 1.5 dex.

\emph{Gadolinium} is concentrated in a spot very similar to that of Nd,
close to the positive magnetic pole region, varying moderately
between $-8.5$ and
$-8.8$\,dex, and is overabundant nearly up to 2.5\,dex relative to the solar value of
$-10.92$\,dex. 

The doubly ionized \emph{terbium} line at 5505\,\AA\ was used to map the surface
distribution of this element. Tb shows a strong overabundance of up to
$-7.7$\,dex (compared to the solar value of $-11.76$\,dex) on the surface
part appearing at the phases where the positive magnetic pole 
dominates the visible stellar hemisphere. It is modulated between $-9.1$ and $-7.7$\,dex,  with the center of maximum abundance 
shifted by $\simeq$ 30\degr\ to higher phases compared to Co, Nd, and Dy.         

\emph{Dysprosium} can also be found in a spot near the positive magnetic pole with the center
of abundance slightly preceding that of e.g. Gd. The element is varying 
between $-8.3$ and
$-8.8$\,dex and found to be overabundant by more than 2.0\,dex relative to the solar value of
$-10.90$\,dex. 

\section{Discussion}             
\label{discuss}

Observing line profiles in several polarization states of stellar radiation 
and modelling their shapes with magnetic 
Doppler imaging, we could derive the magnetic field
geometry and abundance distributions of numerous elements inhomogeneously
distributed over the surface of the roAp
star \hd. This is the first analysis of this kind for a magnetic pulsating star.  
Our investigation shows a complex picture of the interplay 
between magnetic field, pulsation 
and atmospheric inhomogeneities in this prototypical roAp
star. However, the low \vs\ of the star and the limited spectral resolution of our spectropolarimetric
observations did not allow us to reliably recover all but the largest spatial scales of the abundance and magnetic
structures. For this reason we will discuss only the gross properties of
the chemical spots and the magnetic
field, without touching upon the higher-order moments of the surface structures.

We obtain a dominant dipolar magnetic field structure (which might be more complex, 
though the general picture is certainly dipolar) and elemental abundance
patterns that are correlated to this geometry in an unexpected way: 
instead of abundance enhancement regions on \emph{both} magnetic poles or around the magnetic
equator, we observe huge enhancement or depletion regions around \emph{either}
the phase where the positive magnetic pole is visible 
\emph{or} where the magnetic equatorial region dominates the visible stellar surface.

In addition, even more exciting and novel is the fact that these enhancement or depletion
regions for various chemical elements are shifted in longitude relative to each
other.
The center of Mg underabundance as well as that of Sc, Fe, and Ni can be
found at phase 0.0, while that of Ti occurs around phase 0.082. 

The center of maximum abundance of Tb appears very close to phase
$\varphi$\,=\,0.0, that of Co and Gd slightly after $\varphi$\,=\,0.0, around 0.02, 
Nd spots appear at $\varphi$\,=\,0.04, 
Pr and Dy at around $\varphi$\,=\,0.06. Y and La are already shifted 
by 30\degr\ or $\approx$ 0.082 in the direction of increasing phase value.
Ce, whose center of maximum abundance lags behind
all other elements, is shifted by as much as $\approx$\,0.12 or $\approx$\,45\degr.

To investigate the reliability of these longitudinal offsets of the abundance structures we have
performed a forward calculation for Y, La, and Ce showing most of the conspicuous shifts. Using the
final abundance maps reconstructed for these elements we have synthesized their line profiles,
rotating surface structures in the longitudinal direction by small amounts and
compared the resulting modification
of Stokes\,$I$ with the typical observational uncertainty. This comparison is illustrated in Fig.~\ref{Y_La_Ce_rot}
(online material). As is evident from this numerical experiment, the high $S/N$ of our observations
enables us to easily detect shifts of $\ga$\,0.05 for Y and Ce and even smaller shifts for La. Thus the relative longitudinal
positions of spots of different elements are precise to within 0.02--0.05 of the
rotation period. 

\onlfig{9}{
\begin{figure*}[htbp]
\begin{center} 
\vspace{8mm}
\includegraphics[width=150mm]{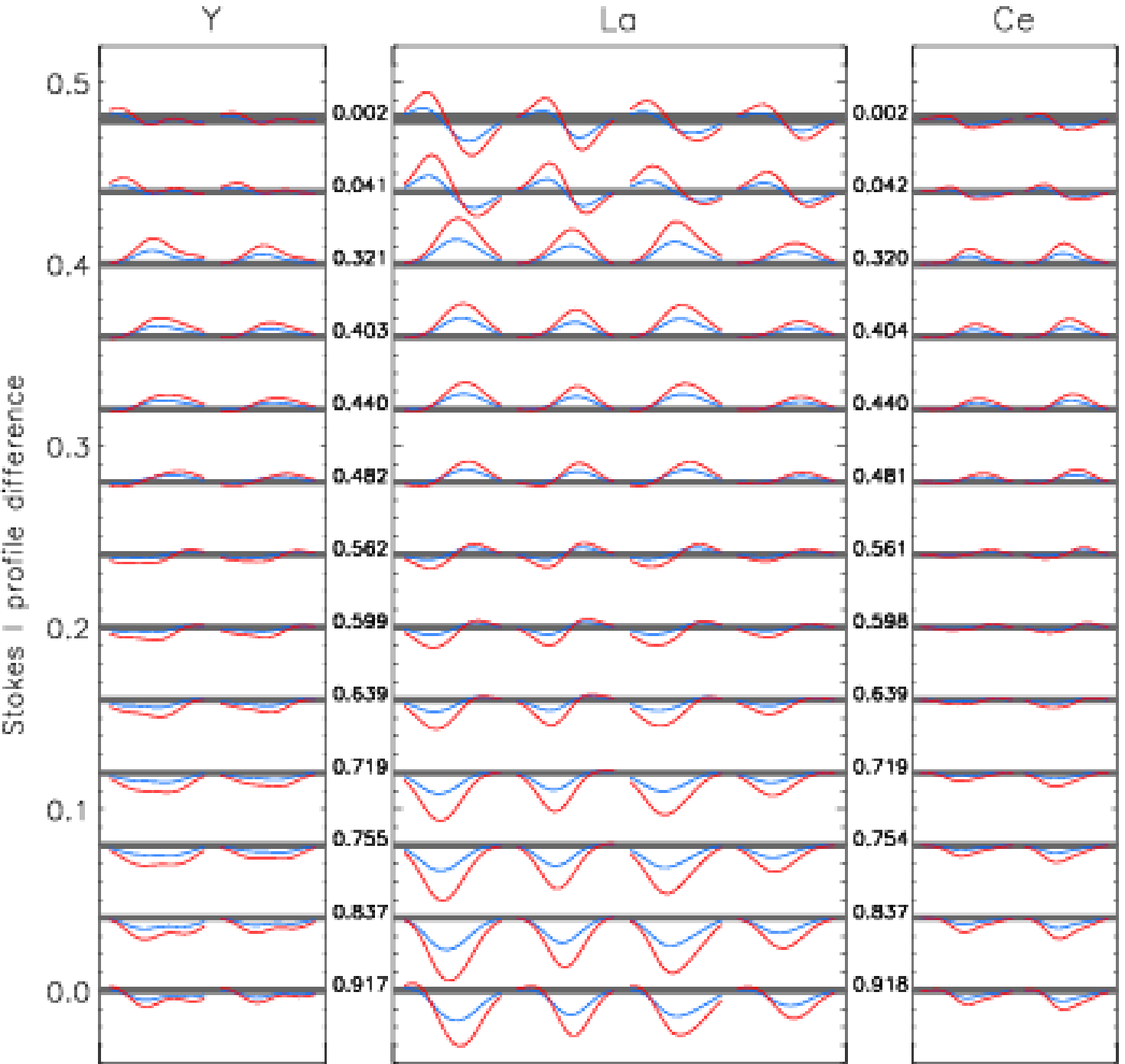}
\caption{Effect of a small rotation of the surface structure of Y, La, and Ce on the Stokes $I$ profiles of the lines
of these elements used for mapping. Panels show the difference between the original synthetic profiles and spectra computed
with the surface spots shifted by 0.05 (blue) and 0.10 (red) of the rotation period (18 and 36\degr\ on the stellar
surface, respectively). The grey bar indicates the uncertainty ($\pm\sigma$) of the spectra corresponding to the $S/N$
of individual
rotation phases. The difference spectra are offset vertically for display
purposes.
The horizontal scale is arbitrary.
}
\label{Y_La_Ce_rot}
\end{center}
\end{figure*}
}

The diverse surface abundance behaviour observed in \hd\ 
most likely gives us information on how the 
magnetic field influences the chemical diffusion of the various species
within the stellar atmosphere. Accordingly we tried to correlate our results with simple
atomic parameters (e.g., atomic weight or 
excitation potential),
but did not find any clear evidence for such a connection. 
This gives rise to the speculation, as the
effect is clearly present, that there is a more complex
correlation of the various atomic characteristics influencing chemical diffusion
 through stellar atmospheres in presence of a magnetic field.  
Thus these relative shifts in longitude of the mapped elements provide valuable
constraints for diffusion models
that try to explain how the magnetic field and other hydrodynamical processes regulate
atmospheric diffusion.

Moreover, we obtain shifts in latitude of abundance enhancement 
regions exhibiting otherwise comparable accumulation tendencies.   
Sc, Ti, Fe,
and Ni in this context appear in similar latitudes, whereas Mg and Co seem
to be shifted to lower latitudes by 30\degr\ and 45\degr, respectively, compared to the 
above mentioned elements.
Among the elements belonging to the group that is enhanced around the positive magnetic
pole, Tb and Y seem to cover the highest latitudes, Ce and Pr are found to be
shifted to lower latitudes by around 15\degr, and La, Nd, Gd, and Dy spots are shifted even more southwards by
$\approx$ 30\degr. 

This finding helps to explain the 
phase shift between pulsation curves of spectral lines belonging to different REE elements, for instance,
Pr and Nd, that we found in the recent study by Ryabchikova~et~al. (\cite{Ryabchik07}). According to the NLTE 
analysis of Nd\,\ii,\iii\ and Pr\,\ii,\iii\ line formation in the atmosphere of HD\,24712 
(Mashonkina~et~al. \cite{nd-NLTE,pr-NLTE}), the lines of both elements are formed at nearly the same 
atmospheric layers. 
However, the pulsation analysis of Ryabchikova~et~al. (\cite{Ryabchik07}) shows that they are shifted by 
0.1 in the pulsation phase relative 
to each other.  According to the theoretical calculations by Sousa\,\&\,Cunha (\cite{SC08}) 
and numerical simulations by Khomenko\,\&\,Kochukhov (\cite{KK09}),
the amplitude and phase 
distributions with height in roAp atmospheres should depend on the local magnetic field strength and
inclination. Slightly different location of Pr and Nd spots relative
to the magnetic field pole
may therefore be a reason for the pulsation amplitude and phase difference observed in Pr and Nd lines. 
This illustrates the potential and importance of MDI when interpreting pulsational observations of
roAp stars.

Vertically stratified abundances are very likely present in the atmosphere
of \hd\ (e.g., Ryabchikova et al. \cite{RKB08}).
In particular, a major discrepancy between two ionization states of the
same REE element is attributed to stratification (Mashonkina et al. \cite{pr-NLTE}). 
To estimate the impact of stratification,
we performed an analysis of the vertical gradient of Fe abundance
(L\"uftinger et al. \cite{Lueftinger04}) during the phase where the magnetic equatorial 
region dominates the stellar hemisphere and also where the positive
magnetic pole is visible, choosing Fe lines that are most sensitive to
stratification. Fe turned out to be significantly stratified, with abundance changing by
$\approx$\,2~dex between the upper and lower parts of the atmosphere of \hd. 
We also found a slight difference in the location of the
abundance jump comparing results from the two extreme phases. In order to keep
MDI results influenced as little as possible by the stratification effects,   
we chose lines for mapping that are hardly affected by a changing vertical
distribution. It is still possible however that part of the horizontal abundance structure
we find is due to variation of the chemical stratification profile across the stellar surface.

Currently, there are no theoretical diffusion models that can provide quantitative explanation of
the abundance maps derived in our Doppler imaging study of \hd. For this reason we are not able to
compare the results  of our investigation to extensive theoretical predictions for individual elements
in the atmospheres of Ap stars. The standard diffusion calculations consider chemical transport
processes in the presence of axisymmetric magnetic field and thus anticipate abundance inhomogeneities
to be axially symmetric themselves (Michaud et al. \cite{MCM81}). These predictions of the simple
diffusion theory are inconsistent with the MDI results obtained for \hd. We find
a nearly dipolar
magnetic field geometry coexisting with diverse abundance distributions, frequently shifted relative to
the magnetic field axis. A similar inconsistency between a simple magnetic field topology and complex
abundance distributions was found in other recent DI studies of Ap stars (Kochukhov et al.
\cite{KPI02,KDP04}). It is possible that diffusion processes are sensitive to the small scale magnetic
structures that could be resolved only with the full Stokes vector MDI, such as the study carried out
by Kochukhov et al. (\cite{KBW04}) for 53~Cam. On the other hand one can also suspect that the current
simple diffusion theory lacks certain physical processes other than the magnetic
field, that can influence
or even drive chemical spot formation. Recent discovery of the evolving chemical clouds in the
atmosphere of \textit{non-magnetic} chemically peculiar stars (Kochukhov et al. \cite{KAG07}) has
demonstrated that the surface structure formation in the upper main sequence stars is not necessarily
driven by the magnetic field alone.

Anticipating further development of theoretical aspects, we provide
new information for the diffusion 
properties within the atmosphere of a cool Ap star, obtain new insights in the atmospheric structure 
and the field geometry, the origin and interplay of abundance variations, pulsation and
magnetic fields in these unique stellar laboratories.

\begin{acknowledgements}
This work was supported by the  Austrian Science Fund (FWF-PP17890) 
and by grant 11630102 from the Royal Swedish Academy of Sciences. 
TR  acknowledges financial support from RFBR grant 
08-02-00469a and from the RAS Presidium (Program ``Origin and Evolution of Stars and Galaxies''). 
O.K. is a Royal Swedish Academy of Sciences Research Fellow supported by a grant from the Knut and Alice Wallenberg Foundation. 
\end{acknowledgements}

\end{document}